\documentclass[dvips,a4paper]{aa}
\usepackage{amsmath}
\usepackage[T1]{fontenc}
\usepackage{epsfig}
\usepackage{natbib}
\usepackage{times}

\bibpunct{(}{)}{;}{a}{}{,}

\newcommand{\sil}{\:\lower0.6ex\hbox{$\stackrel{\textstyle
<}{\sim}$}\:}

\newcommand{\phm}{$\phantom{-}$}

\begin{document}

   \thesaurus{ 06        
              (02.08.1;  
               02.14.1;  
               02.20.1;  
               03.13.4;  
               08.19.4)  
              }

   \title{Thermonuclear explosions of Chandrasekhar-mass C+O white dwarfs}

   \author{M.\ Reinecke     \inst{1}  \and
           W.\ Hillebrandt  \inst{1}  \and
           J.C.\ Niemeyer   \inst{2}
          }

   \offprints{M. Reinecke}

   \institute{Max-Planck-Institut f\"ur Astrophysik,
              Karl-Schwarzschild-Str. 1, 85740 Garching, Germany\
              \and
              LASR, Enrico Fermi Institute, University of Chicago, 
              Chicago, IL 60637, USA\
             }
   \date{Received November dd, 1998; accepted mm dd, yyyy}

   \maketitle

   \begin{abstract}
First results of simulations are presented which compute
the dynamical evolution of a Chandrasekhar-mass white dwarf, consisting
of equal amounts of carbon and oxygen, from the onset of violent
thermonuclear burning, by means of a new two-dimensional numerical code.
Since in the interior of such a massive white dwarf nuclear burning progresses
on microscopic scales as a sharp discontinuity, a so-called flamelet,
which cannot be resolved by any numerical scheme, and since on macroscopic
scales the burning front propagates due to turbulence, we make an attempt
to model both effects explicitly in the framework of a finite-volume 
hydrodynamics code. Turbulence is included by a sub-grid model, following
the spirit of large eddy simulations, and the well-localized burning front
is treated by means of a level set, which allows us to compute the geometrical
structure of the front more accurately than with previous methods.
The only free parameters of our simulations are the location and the amount 
of nuclear fuel that is ignited as an initial perturbation. We find that
models in which explosive carbon burning is ignited at the center
remain bound by the time the front reaches low densities, where we stopped
the computations because our description of combustion is no longer applicable.
In contrast, off-center ignition models give rise to explosions which, however,
are still too weak for typical Type Ia supernovae. Possible reasons for this
rather disappointing result are discussed.

   \end{abstract}

\section{Introduction}
Despite the fact that considerable attention has recently been devoted
to so-called sub-Chandrasekhar mass models (see, e.g.,
\citealt{iben-tutukov-91, woosley-weaver-94, livne-96}),
the most popular progenitor model for at least some Type Ia supernovae 
is a massive white dwarf, consisting of carbon and oxygen,
which approaches the Chandrasekhar mass,
M$_{\text{Ch}}~\simeq~\text{1.39}$~M$_{\odot}$,
by a yet unknown mechanism, presumably accretion from a companion star, and
is disrupted by a thermonuclear explosion (see, e.g., \citealt{woosley-90}
for a review). Arguments in favor of this hypothesis include the ability of 
these models to fit the observed spectra and light curves, in particular
in versions that propagate the burning front mostly in form of a sub-sonic
deflagration wave.

However, not only is the evolution of massive white dwarfs 
to explosion very uncertain, leaving room for some diversity 
in the allowed set of initial conditions (such as the temperature 
profile at ignition), but also the physics of thermonuclear burning
in degenerate matter is complex and not well understood. The generally
accepted scenario is that explosive carbon burning is ignited 
either at the center of the star or off-center in a couple of ignition
spots, depending on the details of the previous evolution. For
example, URCA-neutrino emission affects the temperature
distribution as well as non-steady convection during the slow 
carbon burning phase prior to the explosion, and both effects are 
usually not treated in Type Ia models. After ignition, the flame is
thought to
propagate through the star as a sub-sonic deflagration wave which may or
may not change into a detonation at low densities (around 10$^7$g/cm$^3$).
Numerical models with parameterized velocity of the burning front
have been very successful, the prototype being the W7 model of
\cite{nomoto-etal-84}. However, these models do
not solve the problem because attempts to determine the effective
flame velocity from direct numerical simulations failed and gave
velocities far too low for successful explosions \citep{livne-93,
khokhlov-95, arnett-livne-94a}. This has led to some speculations about ways
to change the deflagration into a supersonic detonation \citep{khokhlov-91a,
khokhlov-91b, khokhlov-91c}.

Here, as well as in several previous papers \citep{niemeyer-hillebrandt-95a,
niemeyer-etal-96, niemeyer-woosley-97},
we aim at a better understanding and a more accurate numerical treatment
of thermonuclear deflagration waves in degenerate C+O white dwarfs, 
hoping to resolve the difficulties of this class of supernova models.
In order to clarify the problem we shall first discuss in Sect. \ref{burnmodel}
the basic physics of deflagration fronts in the flamelet regime
as well as attempts to model them by numerical simulations. In Sect.
\ref {method}
we present a brief description of our new numerical method. The
results of a series of computations are given and discussed in Sect.
\ref{results} and, finally, Sect. \ref{summary} is devoted to a summary
and outlook on future work.

\section{A model for turbulent combustion in the flamelet regime}
\label{burnmodel}
Nuclear burning in the degenerate matter of a dense C+O white dwarf, once 
ignited, is believed to propagate on microscopic scales as a conductive
flame, wrinkled and stretched by local turbulence, but with essentially
the laminar velocity. At high densities near the center of a
Chandrasekhar-mass white dwarf the typical length scales 
for the width of the flame are a fraction of a millimeter.
However, this is not the most relevant length scale. Due to the very
high Reynolds numbers, which are of the order of 10$^{14}$ on typical
macroscopic scales of 10$^7$cm, macroscopic flows are highly turbulent
and interact with the flame,
in principle down to the Kolmogorov scale of 10$^{-3}$cm. This means
that all kinds of hydrodynamic instabilities feed energy into a
turbulent cascade, including the buoyancy-driven Rayleigh-Taylor
instability and the shear-driven Kelvin-Helmholtz instability (see,
e.g., \citealt{niemeyer-hillebrandt-95a} and \citealt{hillebrandt-niemeyer-96}).
Consequently, the picture that emerges is more that of a ``flame
brush'' spread over the entire turbulent regime rather than a wrinkled
flame surface. For such a flame brush, the relevant minimum length
scale is the so-called Gibson scale, defined as the lower bound for
the curvature radius of flame wrinkles caused by turbulent stress.
Thus, if the thermal diffusion scale is much smaller than the Gibson
scale (which is the case for the physical conditions of interest here)
small segments of the flame surface are unaffected by large scale
turbulence and behave as unperturbed laminar flames (``flamelets''). 
On the other hand side, since the Gibson scale is, at high densities,
several orders of magnitude smaller than the integral scale set by
the Rayleigh-Taylor eddies and many orders of magnitude larger than
the thermal diffusion scale, both transport and burning times are
determined by the eddy turnover times, and the effective velocity of
the burning front is independent of the laminar burning velocity.
This general picture has indeed been verified by laboratory combustion
experiments (see, e.g., \citealt{abdelgayed-etal-87}).

\cite{niemeyer-95} and \cite{niemeyer-hillebrandt-95a} have presented a
numerical realization of this general concept. Their basic assumption
was that wherever one finds turbulence this turbulence is fully
developed and homogeneous, i.e. the turbulent velocity fluctuations on
a length scale $l$ are given by the Kolmogorov law 
\begin{equation}
\label{kolmo}
v(l) = v(L) \biggl(\frac{l}{L} \biggr)^{1/3}, 
\end{equation}
where $L$ is the integral scale, assumed to be equal to the
Rayleigh-Taylor scale. Following the ideas outlined above, one can
also assume that the thickness  of the turbulent flame brush on the scale
$l$ is of the order of $l$ itself. With these two assumptions and the
definition of the Gibson scale one finds
 for $l_{\text{gibs}} \sil l \sil L \simeq \lambda _{\text{RT}}$
\begin{equation} 
\label{vtur}
v(l) \simeq u_t(l) \simeq u_t(l_{\text{gibs}}) \biggl(\frac{l}{l_{\text{gibs}}} 
\biggr)^{1/3}  
\end{equation}
and
\begin{equation}
\label{brush}
d_t(l) \simeq l,
\end{equation}
where $v(l_{\text{gibs}}) = u_{\text{lam}}$ defines $l_{\text{gibs}}$,
$u_{\text{lam}}$ is the
laminar burning speed and $u_t(l)$ is the
turbulent flame velocity on the scale $l$.

In a second step this model of turbulent combustion is coupled to a finite
volume hydro scheme such as the PPM-code PROMETHEUS \citep{fryxell-etal-89}.
Since in every finite volume scheme scales smaller than the grid size 
cannot be resolved, we express $l_{\text{gibs}}$ in terms of the grid size
$\Delta$, the (unresolved) turbulent kinetic energy $q$, and the
laminar burning velocity:
\begin{equation}
\label{gibs}
l_{\text{gibs}} = \Delta \biggl( \frac{u_l^2}{2q} \biggr) ^{3/2}.
\end{equation}
Here $q$ is determined from a sub-grid model \citep{clement-93,
niemeyer-hillebrandt-95a} and, finally, the effective turbulent velocity of
the flame brush on scale $\Delta$ is given by 
\begin{equation}
\label{vburn}
u_t(\Delta) = \operatorname{max} ( u_{\text{lam}}, v(\Delta), v_{RT}),
\end{equation}
with $v(\Delta) = \sqrt {2q}$ and $v_{RT} \propto \sqrt {g \Delta}$,
where $g$ is the local gravitational acceleration.

The numerical scheme outlined so far has been applied to two-dimensional
simulations of explosive C-burning in M$_{\text{Ch}}$-mass C+O white dwarfs
by \cite{niemeyer-hillebrandt-95a} and \cite{niemeyer-etal-96}
for a variety of initial conditions. They found that, with the
exception of a model that was ignited off-center in a single small blob, all
models were unbound at the end of the computations, and between 0.59
M$_{\odot}$ (for central ignition) and 0.65 M$_{\odot}$ of matter had been
burnt, which would guarantee an observable, although weak, Type Ia
supernova, but certainly not an average one, in particular because the
expansion velocities were too low.  

There are two possible interpretations of these results. Either they
show that the ``normal'' Type Ia supernova is {\em not} the result of
the explosion of a M$_{\text{Ch}}$-mass C+O white dwarf, caused by a
deflagration wave, or the modeling of thermonuclear combustion
as outlined so far is
still insufficient. In this paper we investigate the second
possibility and, in particular, 
present results obtained by means of an improved numerical
scheme. The main aim is to cure a problem that
is common to all finite volume schemes, namely that, although 
discontinuities are captured automatically, they
are smeared artificially over several grid zones. In many
applications this broadening of discontinuities might not be a big
problem but for modeling turbulent combustion one cannot be satisfied. 

The reasons for this are simple. First, as we discussed already, it is
important to reproduce the geometry of the flame front as accurately
as possible, because its effective surface area also determines the rate of
fuel consumption. If we think of the flame in terms of a surface of 
a given fractal dimension the minimum scale determines essentially its
effective velocity. Therefore it is important to resolve the small
scales as accurately as possible. Secondly, since nuclear reactions are
very sensitive to temperature and their rates depend on a very high
power of $T$, mixing fuel and ashes numerically in certain mesh points
on both sides of the front and smearing out the temperature gradient
there will lead to huge errors in the (local) energy generation
rates. Therefore, if in a code one wants to follow the nuclear
transmutations explicitly one is forced to guarantee that temperature
jumps due to fast reactions are tracked with high precision.

A possible way to reach both goals is to track the front by means of a
level set function and to reconstruct the thermodynamical quantities
in front and behind the the discontinuity from conservation laws.
This approach has been utilized successfully for 
simulations of terrestrial hydrogen
combustion by \cite{smiljanovski-etal-97} and was reformulated
and coded for the application to thermonuclear combustion in C+O white
dwarfs by \cite{reinecke-etal-98}. Here we apply a somewhat simplified
version of this code, i.e. we still assume, as in
\cite{niemeyer-hillebrandt-95a} and \cite{niemeyer-etal-96},
that the matter of the
white dwarf is defined by two states, fuel and ashes,
respectively, and that the transition from fuel to ashes is given by
the instantaneous liberation of a certain amount of energy $q$ once
the material crosses the flame front. Thus the front-tracking by
the level-set function can be interpreted as a kind of passive
advection without reconstruction. Nonetheless, in contrast to common
front capturing schemes, our code is able to resolve complex
structures of the flame on small scales.

  \begin{figure*}
  \centerline{\epsfig{file=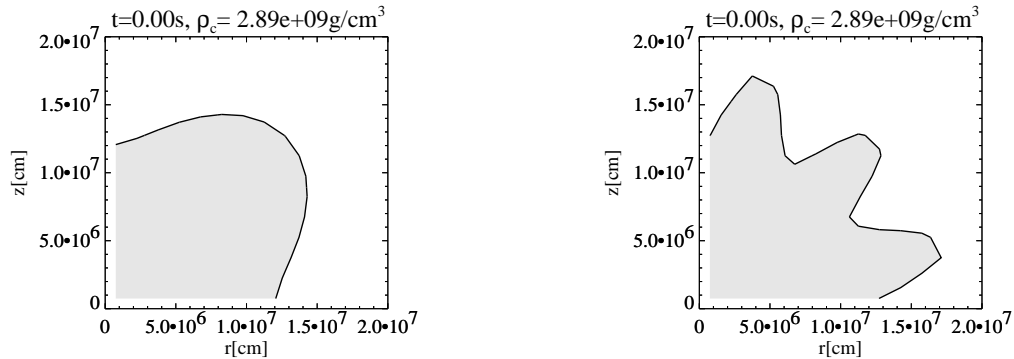, height=0.21\textheight}
    \hspace*{1cm}}
  \caption{Initial front geometry for the centrally ignited models C1 and C3.}
  \label{central-start}
  \end{figure*}

\section{The numerical method}
\label{method}
The numerical code applied in this work is again based on the PPM code
PROMETHEUS \citep{fryxell-etal-89} combined with a sub-grid
model for unresolved turbulence, as was outlined in Sect. \ref{burnmodel}
(see also \cite{niemeyer-hillebrandt-95a}). This includes that the 
propagation velocity of the burning front is obtained from the local 
turbulent kinetic energy as given by eq.(\ref{vburn}).  

As a model for the flame, we use the so-called level set approach first
described by \cite{osher-sethian-88}. The key concept of this model
is to associate the flame front $\Gamma(t)$ with the zero level set of
a scalar function $G(\vec r, t)$:
\begin{equation}
  \Gamma(t):=\lbrace\vec r\ |\ G(\vec{r},t) = 0\rbrace \hspace*{1cm}
  \forall t
\end{equation}
This condition does not fully determine $G$. For numerical reasons it is
appropriate to postulate also that $G$ be positive in the burnt regions
and negative before the front and that
\begin{equation}
 |\vec{\nabla}G| \equiv 1,
\end{equation}
which means that $G$ is a signed distance function.

This flame description is well suited for the combustion in Type Ia supernovae
for a variety of reasons:
\begin{itemize}
\item The front is modeled as a discontinuity between fuel and ashes;
this is an excellent approximation, if one compares the flame witdh and the
typical cell size.
A piecewise linear representation of the front is easy to obtain.
\item Complex topological situations, like the breaking of the burnt area into
several disconnected regions, re-merging and highly convoluted front
geometries are handled without problems.
\item The approach can be used for an arbitrary number of simulated dimensions.
An extension to three-dimensional problems is straightforward.
\end{itemize}

$G(\vec r,0)$ is defined by the initial conditions. During the simulation,
the front propagates with the velocity
\begin{equation}
  \vec v_{\text{flame}}=\vec v_u+u_t\vec n
\end{equation}
with respect to the underlying grid, where $\vec v_u$ represents the velocity
of the unburnt material and $\vec n$ is the front normal pointing towards the
unburnt material. This leads to the following expression
for the temporal evolution of $G$ in the vicinity of the front:
\begin{equation}
      \frac{\partial G}{\partial t}
      = - \vec v_{\text{flame}} \vec{\nabla}G
      = (\vec{v}_u\vec{n} + u_t) |\vec{\nabla}G|
\label{levprop}
\end{equation}
To preserve the distance-function property, an additional re-initialization
step is necessary; see \cite{reinecke-etal-98} and \cite{smiljanovski-etal-97}
for a detailed discussion.

The level set propagation according to eq. \ref{levprop} requires the
knowledge of the velocity of the unburnt material $\vec v_u$ at the front.
Although this value can, in principle, be calculated in the
cells cut by the front \citep{smiljanovski-etal-97}, the extreme
thermodynamical properties of the degenerate white dwarf material make this
reconstruction very difficult.

Fortunately, under the conditions typical for Type Ia supernovae,
the velocity jump is quite small compared to the burning velocity,
so that $\vec v_u$ can be approximated by the average flow velocity with
only small error. In this case, the hydrodynamical advection of $G$ can be
written in conservation form: 
    \begin{equation}
      \int_V \frac{\partial (\rho G)}{\partial t}d^3 r
      + \oint_{\partial V} \vec v_F\rho G d\vec f
      = 0
    \end{equation}
Therefore it is possible to treat the advection of the front caused by the
fluid movement with the Godunov method built into PROMETHEUS.

The propagation of the front, the species conversion and the
energy release due to burning is done in an additional subroutine.
In this step we apply
\begin{equation}
  G'=G+\Delta t u_t \sqrt{D_x^2 + D_y^2},
\end{equation}
where $D_x$ and $D_y$ represent discrete gradients. Next, the burnt volume
fraction of all cells cut by the front are calculated, and the chemical
composition and energy of these cells is adjusted.

\section{Results and discussion}
\label{results}

  \begin{figure*}[p]
  \centerline{\epsfig{file=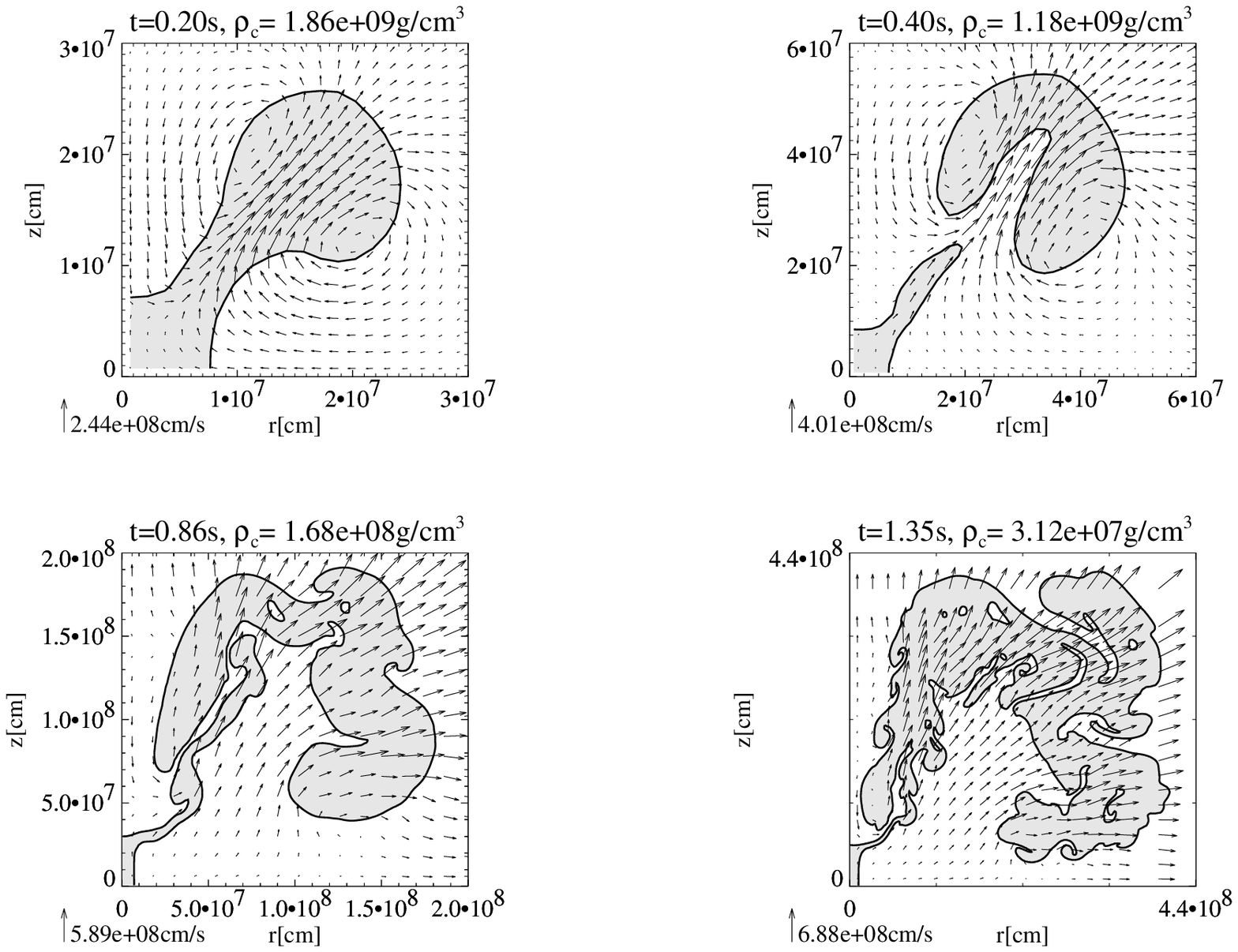, height=0.45\textheight}
    \hspace*{1cm}}
  \caption{Temporal evolution of front geometry and velocity distribution
     for model C1. The inequal spacing of the velocity arrows in the last
     snapshot results from the non-equidistant grid points. Note that
     all scales change from snapshot to snapshot.}
  \label{c1fr}
  \end{figure*}

  \begin{figure*}[p]
  \centerline{\epsfig{file=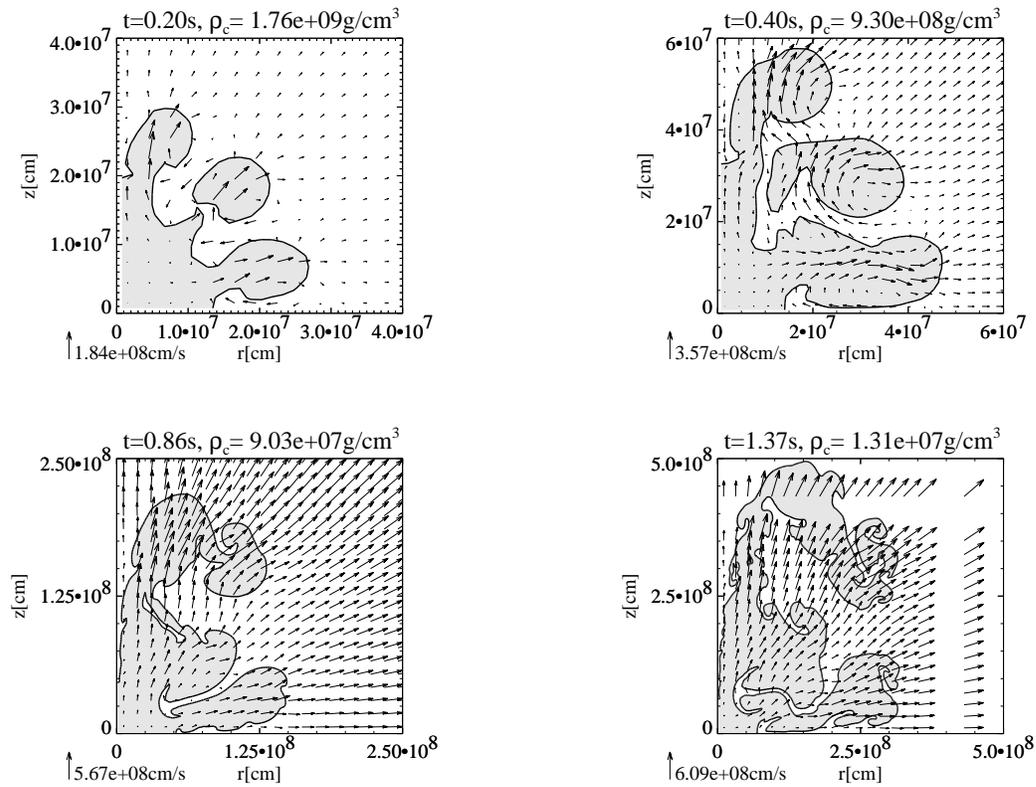, height=0.45\textheight}
    \hspace*{1cm}}
  \caption{Temporal evolution of front geometry and velocity distribution
     for model C3.}
  \label{c3fr}
  \end{figure*}

We have carried out numerical simulations in cylindrical rather 
than in spherical coordinates, mainly because it is much simpler to
implement the level set on a Cartesian (r,z) grid. Moreover, we save
computer time since we avoid very small CFL time-steps which result in
spherical coordinates near the center of the star if one wants to have
good angular resolution.

The grid we used maps the white dwarf onto  256$\times$256 mesh points,
equally spaced for the innermost 226$\times$226 zones by
$\Delta$=1.5$\cdot$10$^6$cm, but increasing by 10\% from zone to zone in
the outer parts. The white dwarf, constructed in hydrostatic
equilibrium for a realistic equation of state, has a central density
of 2.9$\cdot$10$^9$g/cm$^3$, a radius of 1.5$\cdot$10$^8$cm, and a
mass of 2.8$\cdot$10$^{33}$g, identical to the one used by
\cite{niemeyer-hillebrandt-95a}. The initial mass fractions of C and O
are chosen
to be equal, and the total binding energy turns out to be
5.4$\cdot$10$^{50}$erg. Finally, in order to solve for the sub-grid
turbulent kinetic energy, we have to specify its initial values on the
grid and we assume a constant small value of $q$=10$^{12}$erg/g, again in
agreement with \cite{niemeyer-hillebrandt-95a}. This latter assumption
is justified because after a very short time $q$ adjusts to   
the local turbulence, independent of its initial value. At low
densities ($\rho \leq 10^7$g/cm$^3$), the burning velocity of the front is set
equal to zero because the flame enters the distributed regime and our
physical model is no longer valid. However,
since in reality significant amounts of mass may be burned at lower
densities, the energy release obtained from our simulations is a lower
limit only. On the other hand, it is difficult to estimate the error due
to this simplification, mainly because the physics in the distributed regime
of burning is not yet understood. It may happen that the flame just
``stalls'', in which case our numerical description would be appropriate.  
In the other extreme, even a deflagration-to-detonation transition (DDT)
is not excluded which would change the energetics considerably.

  \begin{figure*}
  \centerline{\epsfig{file=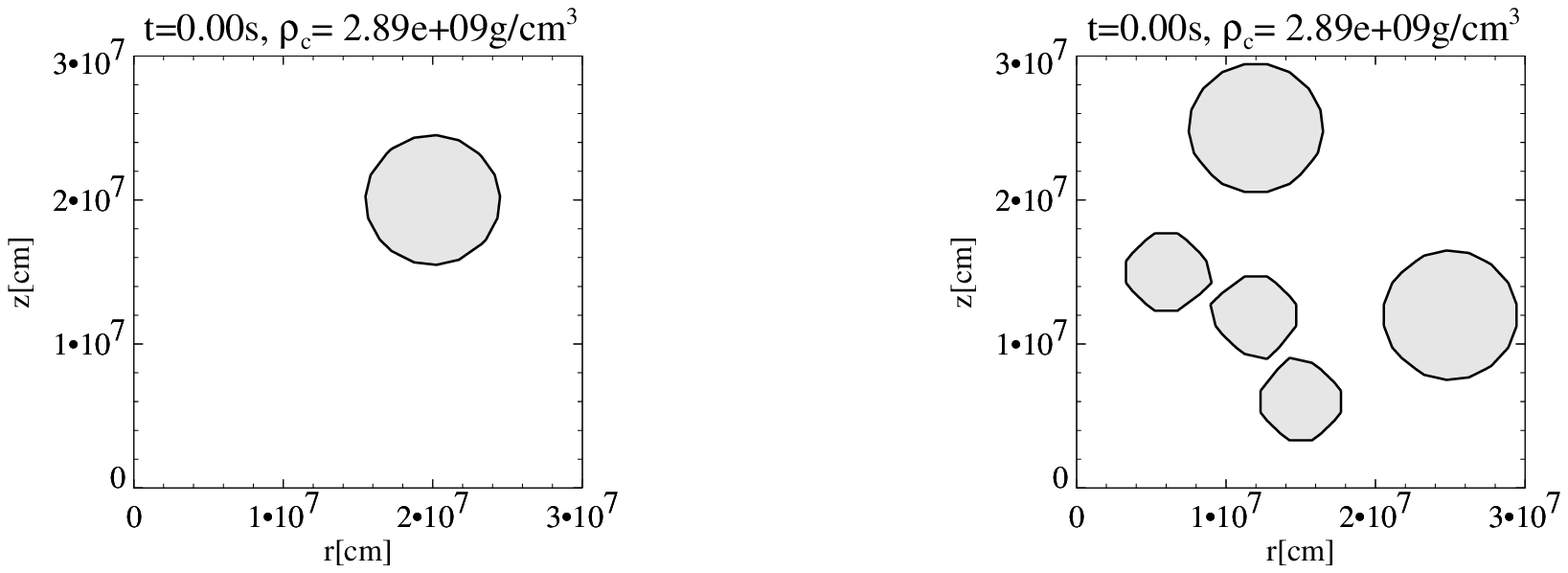, height=0.21\textheight}
    \hspace*{1cm}}
  \caption{Initial front geometry for the off-center ignited models B1 and B5.}
  \label{offcenter-start}
  \end{figure*}

\subsection{Centrally ignited models}

In this set of simulations we start the computations assuming
that the inner 1.5$\cdot$10$^7$cm, corresponding to a mass
of about 0.02M$_{\odot}$, are burnt. In addition, a sinusoidal perturbation
is superimposed producing one ``finger'' (model C1) or three fingers
(model C3), respectively, as seeds for the Rayleigh-Taylor instability
(see Fig. \ref{central-start}).

Fig. \ref{c1fr} shows a sequence of snapshots of the front geometry and the
velocity field for model C1 (note that all scales change 
from snapshot to snapshot). The burnt matter is shaded and the front,
given by the level set function, is plotted as a solid line. 
The evolution follows essentially one's intuition. First the
``finger'' grows into a typical Rayleigh-Taylor structure, mainly in
radial direction, giving rise to two vortices on both sides of the
Rayleigh-Taylor head. There, the turbulence intensity increases and
nuclear burning is accelerated, leading, in turn, to still faster
motions. For a while, a burning blob even disconnects from the much
slower burnt matter near the center of the star. In part, this is due
to our symmetry assumptions which force us to use reflecting boundary
conditions on both coordinate axes, but it also shows that our code can
handle multiply connected fronts.

After about 1s most of the matter of the white dwarf is expanding and
roughly 0.3M$_{\odot}$ have been burnt into Ni, too little  
to unbind the star. The structure of the burnt region becomes very
complex. Kelvin-Helmholtz whirls can be seen at the interfaces between
rising hot matter and unburnt more slowly expanding fuel. Once the
central density has dropped to 3$\cdot$10$^7$g/cm$^3$ we stopped the
calculations for the reason which was explained earlier. The amount of
Ni produced then is nearly 0.33M$_{\odot}$. 
The obvious asymmetries in the flow patterns at late times
are again a consequence of the imposed symmetry conditions:
Convective eddies are not bubbles but axi-symmetric rings, and 
while for an equidistant grid the cell volume does not increase 
in $z$-direction, it does so in radial direction proportional to $r$ 
because of the axi-symmetry.  

Fig. \ref{c3fr} shows snapshots of model C3 where we started the computations
with three ``fingers'' rather than one. The changes again are not
unexpected. Due to the more complex initial perturbation burning
propagates faster in the beginning, mainly because the effective
surface of now three Rayleigh-Taylor bubbles is bigger. The net
effect, however, is that already after about 0.6s the white dwarf
expands and cools, and, in the end, the total amount of Ni produced
is only marginally higher, by about 10\%, as in model C1, but also
this model remains weakly bound.

\subsection{Off-center ignition models}

  \begin{figure*}[p]
  \centerline{\epsfig{file=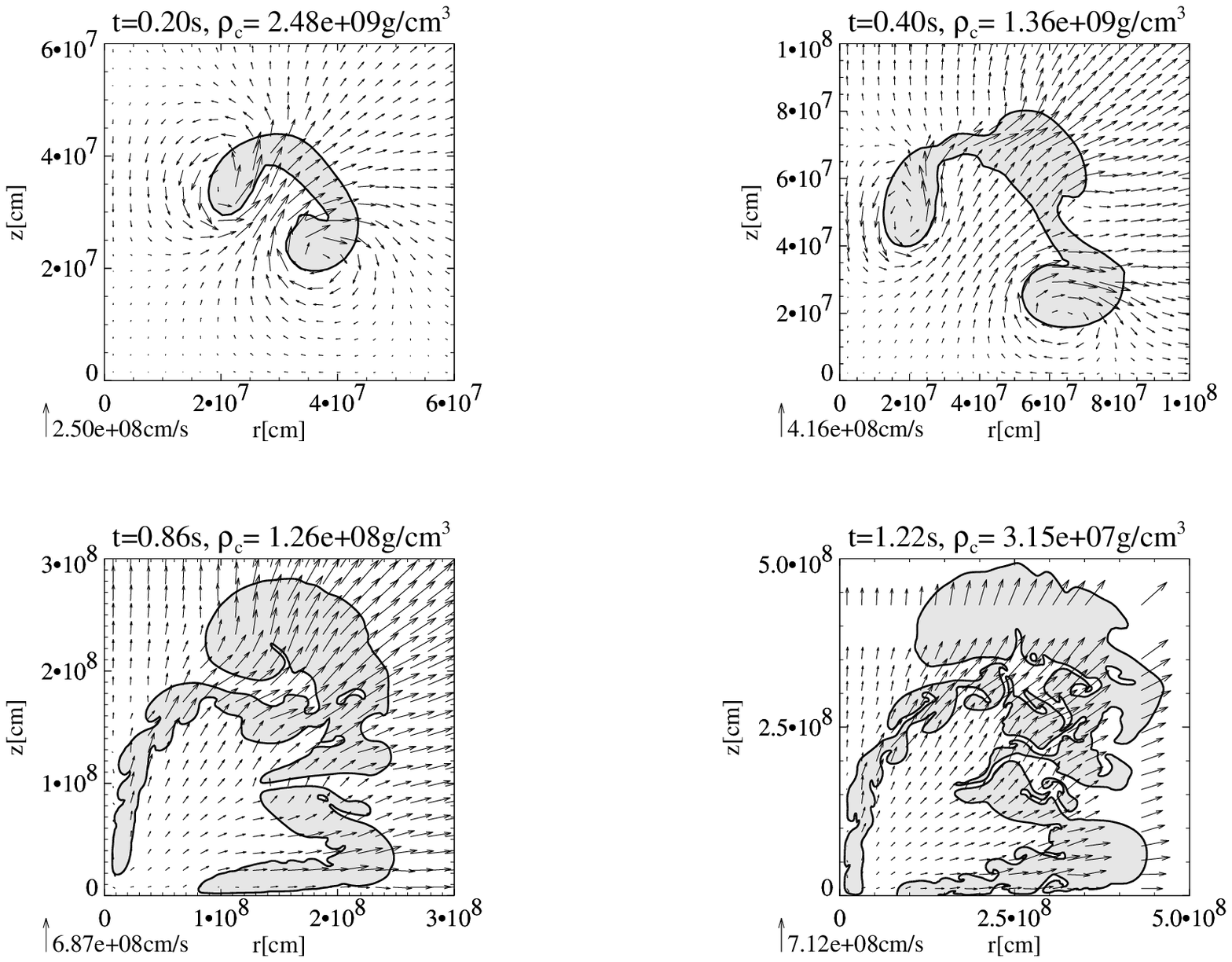, height=0.45\textheight}
    \hspace*{1cm}}
  \caption{Temporal evolution of front geometry and velocity distribution
     for model B1.}
  \label{b1fr}
  \end{figure*}

  \begin{figure*}[p]
  \centerline{\epsfig{file=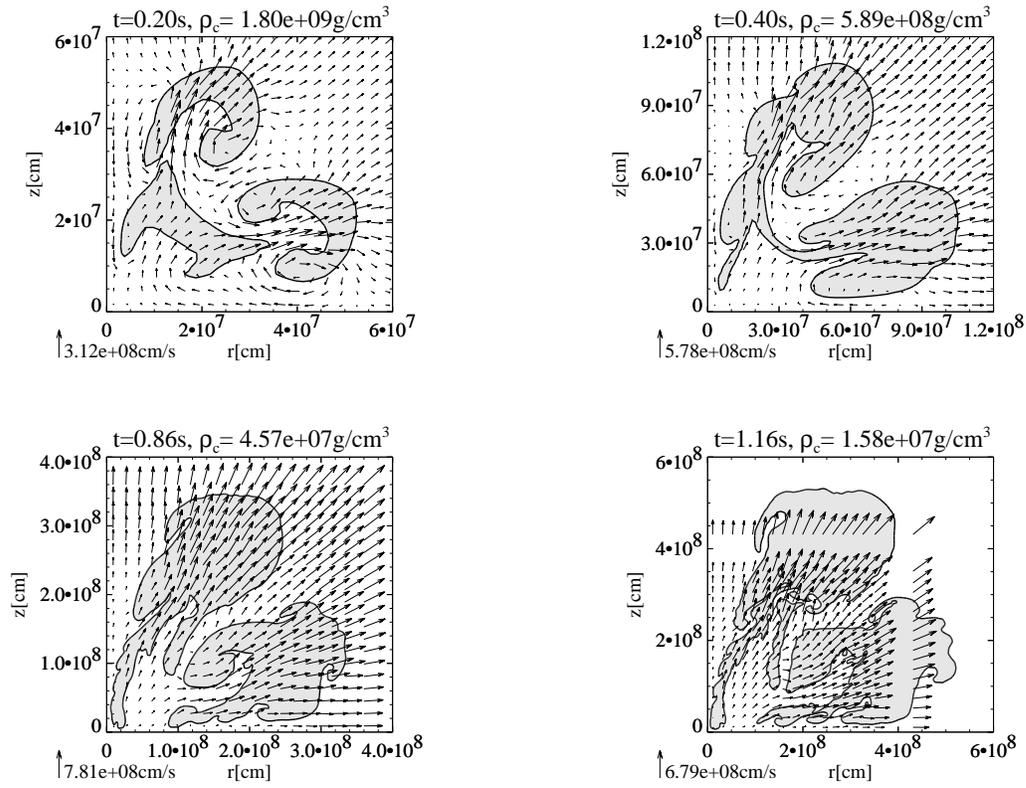, height=0.45\textheight}
    \hspace*{1cm}}
  \caption{Temporal evolution of front geometry and velocity distribution
     for model B5.}
  \label{b5fr}
  \end{figure*}

In two more simulations we follow the work of \cite{niemeyer-etal-96}
and ignite the carbon and oxygen fuel off-center, 
first in one blob and then in five ones (models B1 and B5; see Fig.
\ref{offcenter-start} 
for the initial configurations). Since these blobs can rise freely,
not connected to the center by a symmetry condition, we expect that
they move faster, thereby generating more turbulence and, consequently,
also more rapid burning. Figs. \ref{b1fr} and \ref{b5fr} demonstrate that this
expectation is indeed fulfilled by the simulations. In models B1 
and B5, after 0.61s, peak velocities are typically higher than those 
of models C1 and C3 by up to 50\%. Because
fuel is consumed very rapidly in B1 and B5 also the expansion sets 
in earlier. However, this does not extinguish the burning in the case
of model B5. In contrast, because of the large surface of the various
burning blobs fuel consumption progresses so fast that this model is
unbound already after 0.5s, and continues to burn until the amount of Ni
is 0.34M$_{\odot}$. Therefore, in our computations this model is
the only one which gives rise to an observable supernova
explosion. Model B1, on the other hand side, ends also unbound, but
its total energy at the end of the computations is close to zero.

A final remark concerns numerical effects which are obviously present
in our simulations and which may affect the outcome. First, as was
stated before, in three dimensions without artificial symmetry
assumptions Rayleigh-Taylor eddies will move faster. Moreover, a
multiple of rising blobs has a higher surface area than our
Rayleigh-Taylor ``rings''. Both effects will increase the rate of fuel
consumption and, therefore, will work in favor of more violent
explosions. Moreover, due to finite numerical resolution, the minimum
scale of eddies shown in the previous figures is certainly too large.
For example, whenever blobs disconnect from each other, this is
because burning strings in between them are not resolved. This, in
turn means that our code still underestimates the total rate of 
nuclear burning. Of course, we hope that to a certain extend
the subgrid model for the unresolved turbulent scales takes care
of these effects. However, since we do not resolve the smallest
RT-unstable wavelengths the assumption of a Kolmogorov-spectrum
on sub-grid scales may not be justified and, even more important,
we may not have enough spatial resolution to reach the turbulent
regime. In principle, this hypothesis could be tested by increasing 
the resolution significantly, making the simulations very expensive,
and first results of simulations with $\Delta=5\cdot 10^5$cm give
(after $t\approx0.6$s) indeed a considerably higher energy production 
rate than the models discussed above, indicating that the calculations
are not fully converged. It appears to be unlikely, however, that
going to higher numerical resolution will change our main conclusions.

A numerical comparison of all four models at the end of the simulation is given
in Table \ref{table1}.

\begin{table}
\begin{center}
\begin{tabular}{|l|c|c|c|c|}
\hline
  & C1 & C3 & B1 & B5 \\
\hline \vphantom{\large A}
  Elapsed time [s] & \phm 1.35 & \phm 1.43 & \phm 1.22 & \phm 1.16 \\
\hline \vphantom{\large A}
  Total mass of $^{56}$Ni [M$_\odot$] & \phm 0.23 & \phm 0.26 & \phm 0.27 &
   \phm 0.34 \\
\hline \vphantom{\large A}
  Total energy [10$^{50}$erg] & $-$0.82 & $-$0.23 & \phm 0.04 & \phm 1.32 \\
\hline \vphantom{\large A}
  Kinetic energy [10$^{50}$erg] & \phm 1.95 & \phm 2.00 & \phm 2.55 &
    \phm 3.11 \\
\hline
\end{tabular}
\end{center}
\caption{Comparison of integral quantities at the end of the four simulations.
The numbers given are only approximate because of some mass-loss from
the computational grid at late stages.}
\label{table1}
\end{table}

\subsection{A comparison with previous works}
  \begin{figure*}[p]
    \epsfig{file=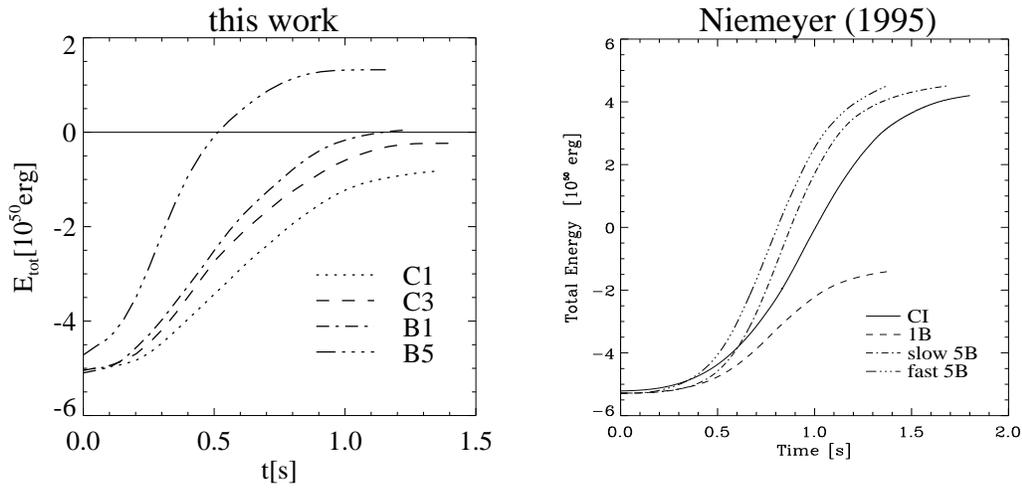,width=0.8\textwidth}
  \caption{Temporal evolution of the total energy}
  \label{etotcomp}
  \end{figure*}
  \begin{figure*}[p]
    \epsfig{file=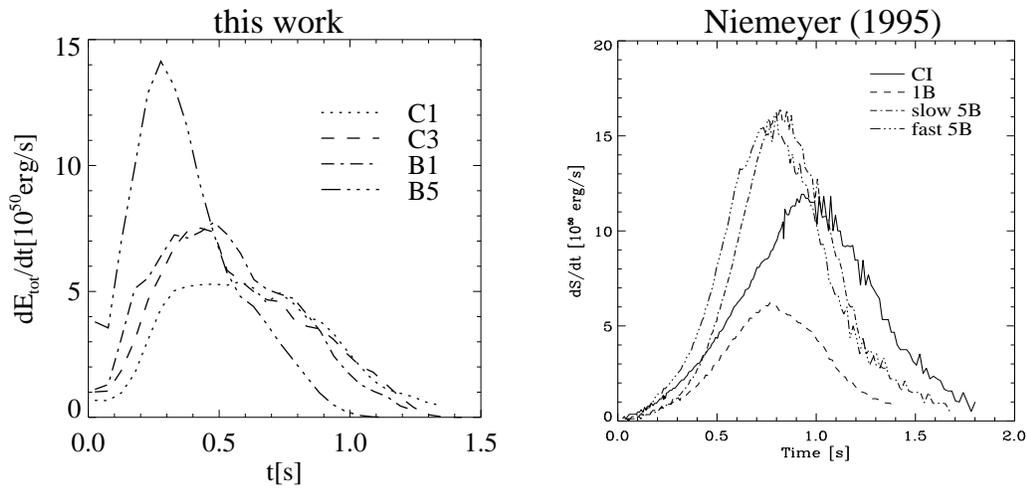,width=0.8\textwidth}
  \caption{Temporal evolution of the energy production by combustion}
  \label{eprodcomp}
  \end{figure*}
  \begin{figure*}[p]
    \epsfig{file=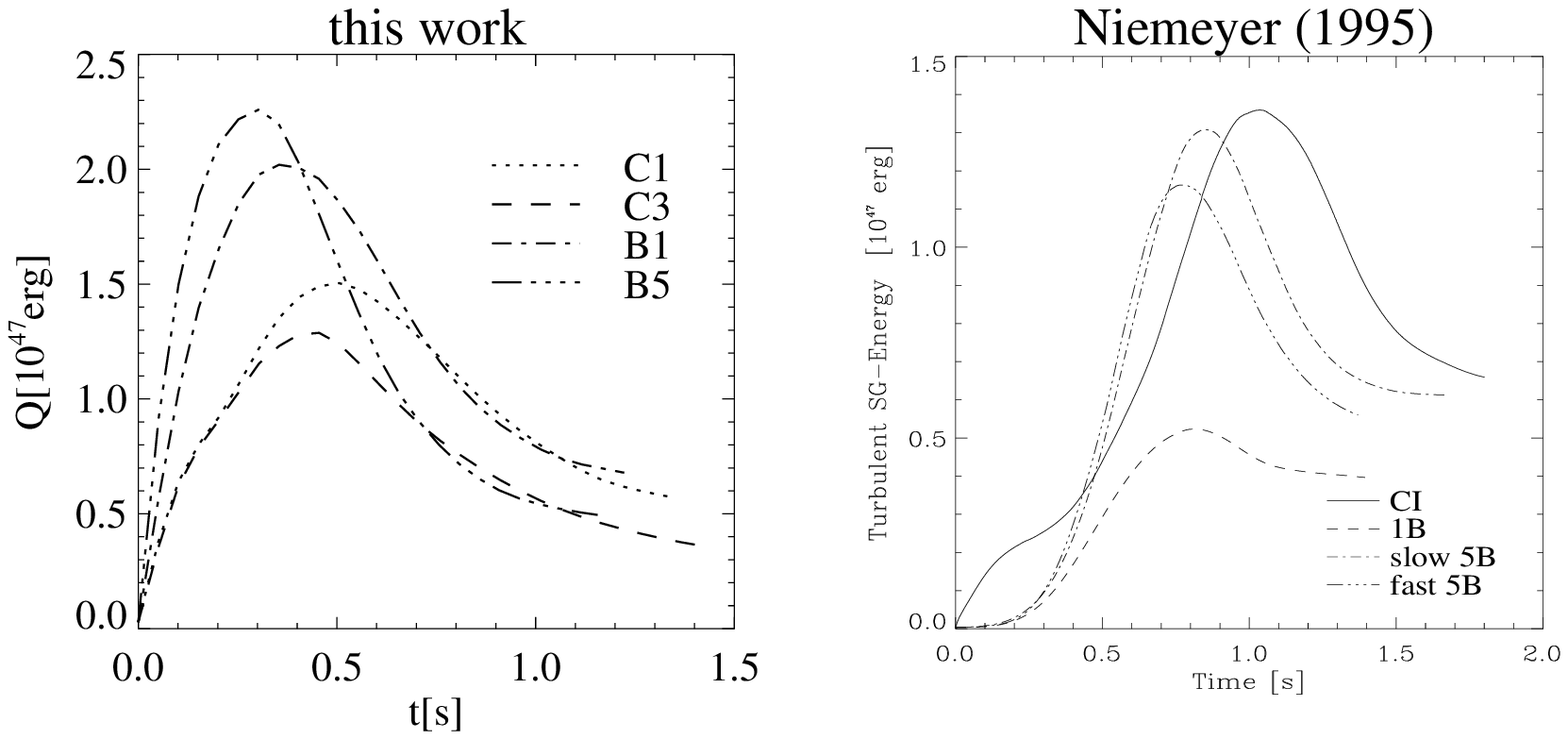,width=0.8\textwidth}
  \caption{Temporal evolution of the turbulent sub-grid energy}
  \label{esubcomp}
  \end{figure*}
The results outlined so far should be compared to those of
\cite{niemeyer-95}, \cite{niemeyer-hillebrandt-95a}
and \cite{niemeyer-etal-96},
because the main motivation behind our present work
was to improve their numerical treatment of a turbulent burning front in
the flamelet regime, hoping that such improvements would change their
weak explosions into more violent ones. Now the contrary appears to be
true. 

In order to make the comparison more quantitative we plot in Figs.
\ref{etotcomp} through \ref{esubcomp}
the total energy, the nuclear energy generation rate, and
the turbulent sub-grid energy as functions of time for our models and
for some models from the previous papers. Since with the exception of
the front-tracking scheme of the present computations both codes are
identical, simulations with similar initial conditions allow us to
compare the effects caused by the flame-propagation model.
In this respect, our model B1 and the model 1B of
\cite{niemeyer-95}, as well as B5 and his ``slow 5B'' are similar.
The same holds for our model C3 and Niemeyer's CI, with the difference that
the perturbation wavelength in our simulation had to be larger due to limited
angular resolution near the center of the star.

The first major difference seen in Fig. \ref{etotcomp} is that, in contrast
to our expectations, the code with front-tracking in general produces 
{\em less} energy from nuclear burning than the other one.
(Note that the energies plotted include already the energy
liberated in the initial perturbation. Therefore the curves do not
start from the same value at time $t$=0s.) But a closer look at
Figs. \ref{eprodcomp} and \ref{esubcomp} reveals the reason. Fig.
\ref{eprodcomp} gives the energy generation
rate for the various models. In those models with front-tracking the
energy generation rate rises more rapidly in the beginning, 
caused by the rapid
increase of the turbulent sub-grid energy (see Fig. \ref{esubcomp}),
in contrast to
the models without front-tracking. This, in fact, is in agreement with
ones expectations because the tracked front has more structure and,
therefore, a larger surface area. However, already after 0.5s
the total energy released
by nuclear burning in all our present models is
high enough to lead to an overall expansion of the white dwarf, as was
discussed in the previous subsections. Consequently, the sub-grid energy
drops again and so does the nuclear energy generation rate. So in
total less energy is produced in comparison to the models without
tracking, which reach their peak values of both the energy generation
and the sub-grid energy about a factor of two later and, therefore,
expand later. An exception is our model B5. Here the energy generation
rate rises so fast that the star is already unbound at the time when
bulk expansion sets in.

\begin{figure}
    \centerline{\epsfig{file=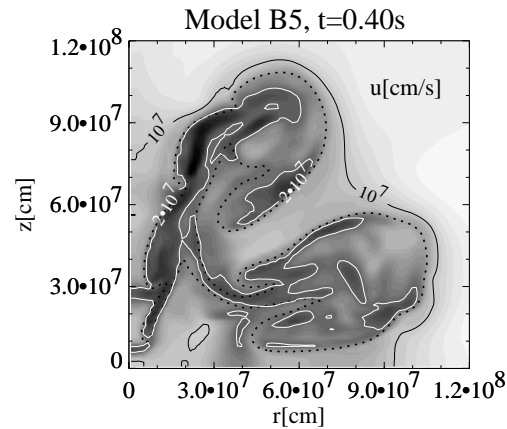,height=0.25\textheight}
    \hspace*{0.5cm}}
  \caption{Distribution of the turbulent velocity fluctuations
    (Model B5, after 0.4s)}
  \label{subspeed}
  \end{figure}

In addition to the higher sub-grid energy production during the early stages
of the simulations, we observe that most of the turbulence is generated in a
very thin region around the flame, resulting in extremely high turbulence
intensities near the front and a further increase in the burning speed
(Fig. \ref{subspeed}). This is caused by the fact that the transition between
the rising bubbles of hot, light ashes and the dense, cold fuel is quite thin
($\approx 2\Delta$), which leads to a well-localized shear flow and turbulent
energy generation, as one would expect in reality.
In Niemeyer's simulations, in contrast, the transition
was smeared out over several cells because of the employed reactive-diffusive
flame model; consequently, the turbulent flame propagation speed was
underestimated.

\section{Summary and conclusions}
\label{summary}
In the present paper we have presented first results obtained by applying a
new numerical technique  to thermonuclear explosions of Chandrasekhar-mass
C+O white dwarfs. Our code differs from those used previously in the
atrophysics literature in that it follows the propagation of the
burning front explicitly by means of a level set function. 
Our approach has the advantage that numerical diffusion of the front 
is largely avoided and that the structure of burning regions
is resolved down to a few grid zones,
which is an important issue in numerical
simulations of combustion. Moreover, in principle,
an extension of the code is possible, making use of the fact that the
thermodynamic properties in mixed cells can be reconstructed from the
conservation laws of mass, energy, and momentum, if the burning velocity
of the front is known. 
Such a code could also be applied to the Type Ia  
supernova problem and would, for the first time, allow 
to include individual nuclear reactions in a meaningful way, and we 
are presently working on this extension.

Since our aim was to demonstrate that deflagrations in M$_{\text{Ch}}$
white dwarfs can lead to explosions which have the properties of some
(if not of typical) Type Ia supernovae, the outcome so far is
disappointing. Our models produce even less nickel and less energy
than those computed with simpler and less accurate numerical schemes. 
In retrospect the reason for this finding is obvious. 
For a ``healthy'' explosion it is not
sufficient to accelerate the burning front beyond what is predicted
by previous numerical experiments, at least not in multi-dimensional
models. In fact, more rapid {\em local} 
burning may under certain circumstances result in an expansion and    
cooling of the white dwarf {\em before} large amounts of nuclear fuel 
are consumed. Our centrally ignited models and the one with 
a single off-center blob are examples.      

The question, therefore, remains whether or no our simulations rule
out the deflagration scenario for typical Type Ia
supernovae. Possible short-comings of the present numerical approach,
such as the artificial symmetry assumption or the still insufficient
numerical resolution, have already been mentioned. Here improvements
will be possible with increasing computer power in the near
future. It might also be worthwhile to investigate the behaviour of
the simulations with varied parameter sets, e.g. for different
chemical compositions or a slightly increased turbulent flame speed,
since both of these quantities are not known exactly.

A second open question concerns our model of turbulent
combustion. Here it is not clear at all, if one of our basic
assumptions, namely that in the presence of reactions the turbulence
spectrum is given by the Kolmogorov law
remains valid. Although in the limit
of very high turbulence intensity experiments seem to support our
hypothesis more work needs to be done. 
Also, significant burning is still possible at low densities in the
so-called distributed regime, and even a transition to a detonation
is not ruled out there \citep{niemeyer-woosley-97}. 

Finally, white dwarfs at the onset of the explosion might look
rather different from the ones we have used as initial
conditions. URCA-neutrino emission and non stationary convection 
during the evolution just prior to the explosion have already been
mentioned as likely sources of large uncertainties. Also, whether the
star reaches the Chandrasekhar-mass by accretion or by merging with a
companion will make a big difference. For example, rotation in some
form may directly affect the propagation of the deflagration front.
All these questions need to be investigated before we can safely dismiss
the deflagration models.

\begin{acknowledgements}
This work was supported in part by the Deutsche Forschungsgemeinschaft under
Grant Hi 534/3-1 and by DOE under contract No. B341495 at the University of
Chicago. The computations were performed at the Rechenzentrum Garching
on a Cray J90.

The authors thank E.\ Bravo for many constructive suggestions which led
to a significant improvement of this paper.
\end{acknowledgements}

\bibliographystyle{aabib}
\bibliography{refs}

\end{document}